\def\astroph{1}
  \ifnum\astroph=0
     \documentclass[manuscript]{aastex}          
  \else
    \documentclass{emulateapj}
   \fi

  \usepackage{epsfig,amsmath,natbib}
  \usepackage[dvips]{color}
  \usepackage{verbatim}

\begin{document}
\shorttitle{Ly$\alpha$ Resonant Scattering in Young Galaxies}
\shortauthors{Laursen et al.}
\title{Lyman {\Large $\alpha$} Resonant Scattering in Young Galaxies --- Predictions
from Cosmological Simulations}
\author{Peter Laursen\altaffilmark{1}
        and Jesper Sommer-Larsen\altaffilmark{1,2}}
\email{pela@dark-cosmology.dk, jslarsen@tac.dk}
\altaffiltext{1}{Dark Cosmology Centre, Niels Bohr Intitute, University of
                 Copenhagen, Juliane Maries Vej~30, DK-2100, Copenhagen {\O},
                 Denmark.}
\altaffiltext{2}{Institute of Astronomy, University of Tokyo, Osawa 2-21-1,
Mitaka, Tokyo, 181-0015, Japan.}
\begin{abstract}
We present results obtained with a three-dimensional, Ly$\alpha$ radiative
transfer code applied to a fully cosmological galaxy formation
simulation. The developed Monte Carlo code is capable
of treating an arbitrary distribution of source Ly$\alpha$ emission, neutral
hydrogen density, temperature, and peculiar velocity of the interstellar
medium.
We investigate the influence of resonant scattering on the appearance and
properties of young galaxies by applying the code to a simulated ``Lyman-break
galaxy'' at redshift 
$z = 3.6$, and of star formation rate 22 $M_{\odot}$ yr$^{-1}$ and total 
Ly$\alpha$ luminosity $2.0\times10^{43}$ erg s$^{-1}$. It is found that
resonant scattering of Ly$\alpha$ 
radiation can explain that young galaxies are frequently observed to be more
extended on the sky in Ly$\alpha$ than in the optical. Moreover, it is
shown that, for the system investigated, due to the anisotropic escape of the 
photons, the observed maximum surface brightness can differ by more than an
order of magnitude, and the total
derived luminosity by a factor of $\sim 4$, depending on the orientation of
the system relative to the observer.
\end{abstract}

\section{Introduction}
The Ly$\alpha$ line is a very important diagnostic in a wide range of fields
of astrophysics, not the least of which is galaxy formation, providing us with
extensive
information on redshift, dynamics, kinematics, morphology, etc. Three 
distinct physical processes result in Ly$\alpha$ source emission in the 
context of galaxies:
First, Ly$\alpha$ emission due to photo-ionization 
of hydrogen atoms by UV radiation from nearby, massive stars
and subsequent recombinations may contribute as much as 10\% of the 
total luminosity of the galaxy \citep{par67}.
Second, part of the potential energy gained by gas falling into galactic
potential wells is converted into cooling radiation. \citet*{far01} find 
that, at high redshifts, most of this radiation is emitted by gas with 
$T < 20\,000$~K, and consequently $\sim 50\%$ in Ly$\alpha$ alone.
Finally, the external, metagalactic UV field, penetrating some (in case of 
damped
Ly$\alpha$ systems) or all (in case of Lyman-limit systems) of the
outer parts of galactic hydrogen ``envelopes'' will also produce some
Ly$\alpha$ radiation through case B
recombination.
Moreover, this UV field can also photo-heat non-self-shielded gas,
which subsequently cools, radiating Ly$\alpha$ \citep{fur05}.

Over the past years it has become possible to actually resolve
observationally these young, Ly$\alpha$ emitting galaxies. In several cases,
the galaxies have been
found to be significantly more extended on the sky when observed in Ly$\alpha$
as opposed to optical bands \citep[e.g.,][]{mol98,fyn01,fyn03}.
Due to the complexity and
diversity of the systems, in order to correctly interpret observations and 
make predictions about the properties of young galaxies, it is desirable to
develop realistic theoretical and numerical models.

For a number of idealized cases analytical solution are obtainable.
\citet{har73} investigated the emergent spectrum of resonantly scattered 
radiation in the case of a highly optically thick slab of 
finite thickness but infinite extension, and 
uniform temperature and density.
\citet{neu90} extended this solution to include the possibility of the photons
being destroyed (as by dust) and injected with arbitrary initial frequency.
\citet{dij06} derived a similar analytical expression for spherical symmetry,
allowing for isotropic expansion or collapse of the gas, and
\citet{loe99} examined the spectrum for an isotropically expanding or
contracting medium with no thermal motion.

However idealized these configurations may seem, they provide valuable and
at least qualitative insight into the characteristics of young galaxies.
Moreover, they offer direct means of testing numerical methods.

The Monte Carlo (MC) method 
has been used for solving radiative transfer (RT) problems since the
beginning of the 1960s.
Nevertheless, though conceptually simple, the demand for strong
computer power until quite recently restricted this technique to deal with
more or less the same idealized configurations that had already been dealt with
analytically. Thus, the majority of previous attempts to
model RT in astrophysical situations have been based on strongly simplified
configurations of the physical parameters. Only a few authors
\citep{can05,tas06,ver06} considered more general cases.

With the aim of predicting the appearance and properties of
Ly$\alpha$ emitting galaxies, we developed an MC code capable of treating an
arbitrary distribution of source Ly$\alpha$ emission, neutral
hydrogen density, temperature, and peculiar velocity of the medium and
subsequently applied it to a simulated Lyman-break galaxy (LBG) from a fully
cosmological simulation.
\ \\
\ \\

\section{Simulations}
\label{sec:code}

\subsection{The Cosmological Simulation and Monte Carlo Code}
\label{sec:ini}

The cosmological simulation of the formation and evolution of an individual
galaxy was performed using the N-body/hydrodynamical 
TreeSPH code of \citet[][see also \citet{som06}]{som03}.
The system becomes a Milky Way/M31-like disk galaxy at $z = 0$ and is
simulated using $\sim2.2\times10^6$ particles in total, comprising only
smoothed particle hydrodynamic (SPH) and dark matter (DM) particles at the
initial redshift $z_i = 39$. 
The masses and gravity softening lengths
of SPH and star particles are $9.9 \times 10^4 h^{-1} M_{\odot}$ and
$200 h^{-1}$ pc, respectively ($h = 0.65$). For the DM particles the
corresponding
values are $5.7 \times 10^5 h^{-1} M_{\odot}$ and $370 h^{-1}$ pc. The minimum
SPH smoothing length in the simulation is about $13 h^{-1}$ pc.
For the purposes of the present study, a single simulation output, at
$z = 3.6$, was chosen.

The MC code, used to propagate the photons through the medium, by and
large resembles those recently developed by other authors
\citep[e.g.,][]{can05,tas06}. Since it is grid-based --- the number of cells
typically being 512$^{3}$ --- the physical parameters of interest are first
interpolated from the
SPH particles to
the cells of the computational box. These parameters are the Ly$\alpha$
emissivity, the temperature $T$, the density
$n_{\textrm{{\scriptsize H}{\tiny I}}}$ of neutral hydrogen, and
the three-dimensional velocity field $\mathbf{v}_{\mathrm{bulk}}$ of the gas.

\subsection{Determination of Source Ly$\alpha$ Emission}
\label{sec:em}

The emission cell of a given Ly$\alpha$ photon is found by setting the
probability of being emitted from a given cell equal to the ratio of the
luminosity in that cell to the total luminosity $L_{\mathrm{tot}}$. The photon
is then injected in the line center (in the reference frame of the fluid
element) from a random point $\mathbf{x}_i$ in the
cell.
The frequency $\nu$ of the photon is parametrized through
$x = (\nu - \nu_0)/\Delta\nu_{\mathrm{D}}$, where $\nu_0 = 2.466\times10^{15}$
Hz is the line center frequency and
$\Delta\nu_{\mathrm{D}} = (v_{\mathrm{th}}/c)\nu_0$ is the Doppler frequency,
with $v_{\mathrm{th}} = (2 k_B T / m_{\mathrm{H}})^{1/2}$ being the thermal 
atom velocity dispersion (times $\sqrt{2}$) and the rest of the variables
having their usual meanings. In terms of this variable the injected photon
obviously has a frequency of $x = 0$. 
The initial direction
$\mathbf{\hat{n}}_i$ of the photon follows an isotropic probability
distribution.

\subsection{Propagation of the Radiation}
\label{sec:prop}

The optical depth $\tau$ covered by the photon before it is scattered is 
drawn randomly from the probability distribution $e^{-\tau}$, and subsequently
converted into a physical distance
$r = \tau / n_{\textrm{{\scriptsize H}{\tiny I}}} \sigma_x$, where
the physical parameters are given by the present cell. The
cross section $\sigma_x$ is given by a Voigt profile, i.e.~the convolution of
the Lorentzian natural line profile and the Gaussian thermal broadening of the
atoms, resulting in
\begin{equation}
\label{eq:xsec}
\sigma_x = f_{12} \frac{\sqrt{\pi} e^2}{m_e c \Delta\nu_{\mathrm{D}}} H(a,x),
\end{equation}
where $f_{12} = 0.4162$ is the Ly$\alpha$ oscillator strength, and
\begin{equation}
\label{eq:H}
H(a,x) = \frac{a}{\pi} \int_{-\infty}^{+\infty}
         \frac{e^{-y^2}}{(x-y)^2 + a^2} dy
\end{equation}
is the Voigt function with $a = \Delta\nu_{\mathrm{L}}/2\Delta\nu_{\mathrm{D}}$
the ratio between the natural line width 
$\Delta\nu_{\mathrm{L}} = 9.936\times10^{7}$ Hz and the Doppler width.
Since eq.~(\ref{eq:H}) is not analytically
integrable, we use the analytic fit given by \citet{tas06}.

If the final position $\mathbf{x}_f$ of the photon is outside the original
cell, the photon is placed at the point $\mathbf{x}_{\mathrm{int}}$
of intersection with the face of the
cell and the above calculation is redone with the parameters of the new cell,
an optical depth $\tau' = \tau - |\mathbf{x}_{\mathrm{int}} - \mathbf{x}_i|\,
( n_{\textrm{{\scriptsize H}{\tiny I}}}\,
\sigma_x )_{\mathrm{prev.cell}}$, and a frequency Lorentz transformed to the
bulk velocity of the new cell. This procedure is repeated until either the
originally assigned value of $\tau$ is spent, and the photon is scattered, or
it leaves the computational box.

\subsection{Scattering}
\label{sec:scat}

Except for a small recoil effect, the scattering is coherent in the reference
frame of the atom. However, to an external observer, the non-zero velocity
$\mathbf{u} = \mathbf{v}_{\mathrm{atom}}/v_{\mathrm{th}}$ of
the atom will shift the frequency of the photon. In the directions
perpendicular to $\mathbf{\hat{n}}_i$,
the velocities $u_{\perp1,2}$ will follow a Gaussian
distribution. When $x \sim 0$, the photon barely diffuses spatially. Only when
it has diffused sufficiently far in frequency space, will it be able to make
a large journey in real space. To skip these non-important core scatterings and
thus accelerate the code, if $|x|$ is less than some critical value
$x_{\mathrm{crit}}$, following \citet{dij06} $u_{\perp1,2}$ is drawn from a
truncated Gaussian so as
to favor fast moving atoms and artificially push the photon back in the wing.
$x_{\mathrm{crit}}$ is determined according to $a \tau_0$ in the given cell. If
$a \tau_0 \le 1$, a proper Gaussian is used; otherwise we find that
$x_{\mathrm{crit}} = 0.02\,e^{\xi \ln^{\chi} a\tau_0}$, where
$(\xi,\chi) = (0.6,1.2)$ or $(1.4,0.6)$ for
$a\tau_0 \le 60$ or $a\tau_0 > 60$, respectively, can be used without
affecting the final result. The effect of this acceleration scheme is a
speed-up of several orders of magnitude.

Due to the resonance nature of the scattering event,
the velocity $u_{||}$ parallel to $\mathbf{\hat{n}}_i$ depends on $x$, and is
generated following \citet{zhe02}.

The final frequency $x'$ of the scattered photon (in the reference frame of the
fluid element)
depends on direction in which the photon is scattered. For scattering in the 
line center, transitions to the $2P_{1/2}$ state results in
isotropic scattering, while the $2P_{3/2}$ transition causes some polarization,
resulting in a probability distribution
$W(\theta) \propto 1 + \frac{3}{7}\cos^2\theta$, where $\theta$ is the angle
between $\mathbf{\hat{n}}_i$ and the outgoing direction $\mathbf{\hat{n}}_f
$\citep{ham40}. Since the spin multiplicity is
$2J + 1$, the probability of being excited to the $2P_{3/2}$ state is twice as
large as being excited to the $2P_{1/2}$ state\footnote{For the environments 
produced here, transitions to the $2S$ state and subsequent
destruction of the photon can be neglected.}. For scatterings in the wing,
polarization for $\pi/2$ scattering is maximal, resulting in a dipole 
distribution $W(\theta) \propto 1 + \cos^2\theta$ \citep{ste80}. The
transition from core to wing scattering is given by the value of $x$ where
the Lorentzian starts dominating the Gaussian profile, i.e.~where
$a/\pi x^2 \sim \sqrt{\pi} e^{-x^2}$, or $x \sim 3$ (the exact value is not
crucial). In all
cases, the scattering is isotropic in the azimuthal angle $\phi$. In the
observers frame, the final frequency is then given by
$x' = x - u_{||} + \mathbf{\hat{n}}_f \cdot \mathbf{u}
+ g(1 - \mathbf{\hat{n}}_i \cdot \mathbf{\hat{n}}_f)$,
where the factor $g = h_{\mathrm{Pl}} \nu_0 / m_{\mathrm{H}} c v_{\mathrm{th}}$
\citep{fie59} accounts for the recoil effect.

\subsection{Observable Surface Brightness Maps}
\label{sec:obs}

Following the above scheme, the photons are propagated through the medium, one
by one, until they escape the computational box. For each scattering
(as well as the emission) of a photon, the probability that the photon will
escape in the direction of
six virtual observers situated in the negative and positive directions of the
three principal axes is calculated. This probability is added as a weight to a
three
dimensional array (the two spatial dimensions of the projected image of the
galaxy, plus a spectral dimension for each pixel). The array is finally
collapsed along the spectral dimension and along the spatial dimensions to 
yield the image and spectrum, respectively, that an external observer would
see. 
The contribution of each pixel element to the surface
brightness (SB) is then
%
\begin{equation}
\label{eq:SB}
\mathrm{SB_{\mathrm{pix}}} = \frac{L_{\mathrm{tot}}}{n_{\mathrm{ph}}}
                             \frac{1}{d_L^2}
                             \frac{1}{\Omega_{\mathrm{pix}}}
                             \sum_{\mathrm{ph.,scat.}} 
                             e^{-\tau_{\mathrm{esc}}} W(\theta),
\end{equation}
where $n_{\mathrm{ph}}$ is the number of photons, $d_L$ is the luminosity
distance to the observer, $\Omega_{\mathrm{pix}}$ is the solid angle subtended
by the pixel, and $\tau_{\mathrm{esc}}$ is the optical depth of the gas lying
between the scattering event and the edge of the computational box in the
direction of the observer.\\[.3cm]
The MC code was tested on various simple configurations for which analytic
solutions exist, discussed in the introduction. Our code exquisitely passes all
tests. The results of these tests will be presented in a future paper.

\section{Results}
\label{sec:res}
The MC code was applied to a proto-galaxy at $z = 3.6$, consisting of two small,
star-forming ``disks'' separated by approximately 2 kpc, on one of which the
computational box is centered, and a third more extended disk of lower star
formation rate (SFR),
located about 15 kpc from the center. The star-forming
regions are embedded in a significant amount of more diffuse, non-star-forming
H\textsc{i} gas in a 10--15 kpc
thick, sheet-like structure, taken to constitute the $x$-$y$ plane.
The total SFR of the system is 22 $M_\odot$ yr$^{-1}$. Observed
LBGs have SFRs in the range
10--1000 $M_\odot$ yr$^{-1}$ \citep[e.g.,][]{rig06}, so the
simulated galaxy corresponds to a small LBG.
The Ly$\alpha$ emissivity is produced according the three different scenarios
described in the introduction. Specifically, the luminosity originating from
the H\textsc{ii} in the vicinity of massive stars (accounting for
approximately 90\% of $L_{\mathrm{tot}}$) is determined following
\citet{far01}, using the code Starburst99 \citep{lei99} to yield the Lyman
continuum and
assuming a Miller-Scalo initial mass function,
a mean Lyman continuum photon energy of 1.4
Rydberg, and that 0.68 Ly$\alpha$ photons are emitted per photonionization.
The length of the box used for the MC calculations is 50 kpc (physical),
and $d_L \sim 34$ Gpc.

\begin{figure}
\epsscale{1.2}
\plotone{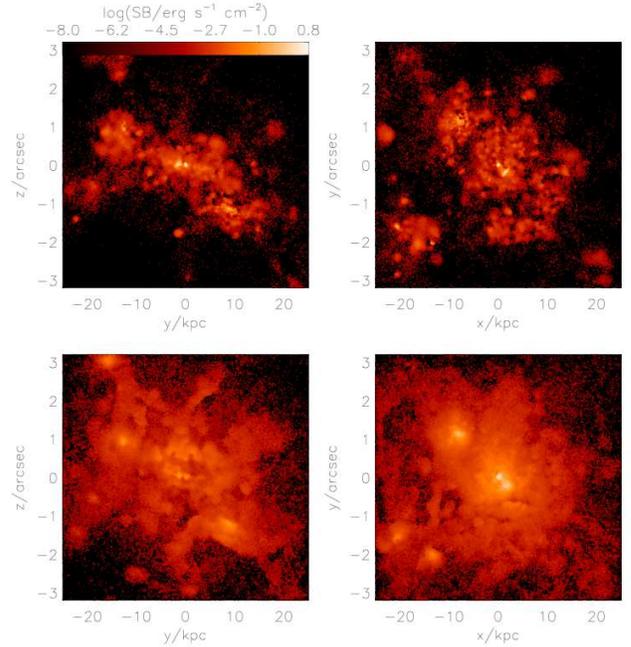}
\caption{Bolometric surface brightness map of a simulated galaxy at $z = 3.6$.
         Left and right column shows the system when viewed edge-on
         and face-on, respectively. The top panel displays the galaxy as if
         the Ly$\alpha$ radiation was able to escape directly, without
         scattering. The bottom panel shows the effect of the scattering.}
\label{fig:SBmap}
\end{figure}

Fig.~\ref{fig:SBmap} shows the results obtained (all with a resolution of
$512^3$ cells), viewed from two different 
directions --- from the negative $x$- and $z$-direction, 
corresponding to an edge-on and a face-on view of the sheet-like structure, 
respectively. Upper panels assume that the gas is optically thin to the
Ly$\alpha$ line, lower show the corresponding results with resonant 
scattering included.

\begin{figure*}
\epsscale{1}
\plotone{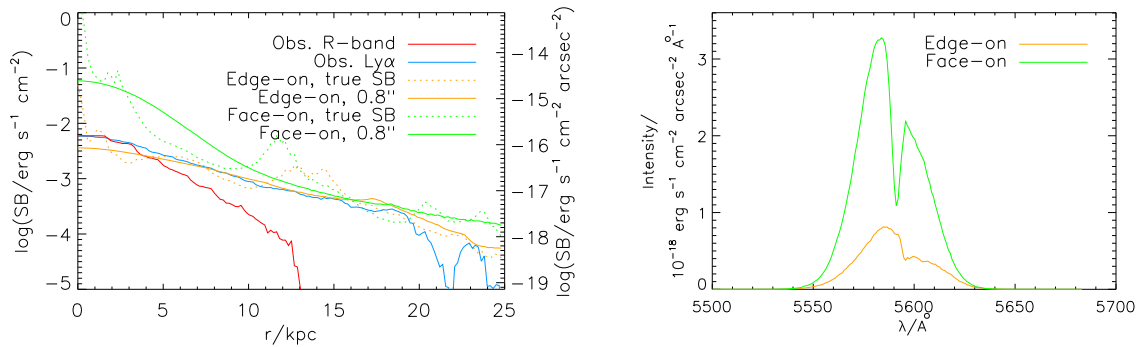}
\caption{\emph{Left}: Bolometric surface brightness (SB) profiles of the galaxy
         when viewed edge-on (\emph{orange}) and face-on (\emph{green}). Both the
         true (\emph{dotted}) and the smoothed (\emph{solid}, see text) profiles are
         displayed.
         Also shown are the SB profiles of the galaxy LEGO2138\_29 \citep[SFR
         $\sim 15$ $M_\odot$ yr$^{-1}$ if extinction is negligible,][]{fyn03} in
         Ly$\alpha$ (\emph{blue}) and in the $R$-band (\emph{red}, normalized to the maximum
         observed Ly$\alpha$ SB). In particular the SB of the $y$-$z$ plane
         nicely reproduces the observed SB.
         Left axes measure the SB at the source, while right axes measure the
         SB observed at $z = 0$.
         \emph{Right}: Spectrum of the emergent radiation, clearly displaying
         the characteristic double-peaked profile. The enhanced blue peak
         indicates a net inward bulk velocity of the gas.}
\label{fig:ProfSpec}
\end{figure*}

The effect of the scattering is incontestable: although the original
constellation of the principal emitters is still visible, the surface
brightness distribution is clearly much more extended. Moreover, 
we notice the effect of the viewing angle. Qualitatively, we expect the 
photons to escape more easily
from the face of the sheet than from the edge and hence that the system should
have a higher surface brightness than when viewed edge-on. Here, this
anticipation is quantified: Fig.~\ref{fig:ProfSpec} shows the azimuthally
averaged SB profiles.
To allow for direct comparison with observations, the profiles are also
shown not including
the luminosity of the emitter lying in the outskirts of the image and
smoothed with a PSF corresponding to a seeing of $0.8''$. The
maximum SB of the $x$-$y$ plane
is found to be $6.1 \times 10^{-2}$ erg s$^{-1}$ cm$^{-2}$, or $\sim14$ times
higher than that of the $y$-$z$ plane. The average
SB, from which the total luminosity is usually derived assuming isotropic
emission, is $3.8$ times higher.

Finally, we show the emergent spectrum. Though not as clear for the edge-on
view, both profiles exhibit the characteristic double-peaked profile
\citep[see, e.g.,][]{ven05}.
Obviously, this is the result of the high opacity for photons near the line
center; diffusing to either side of $\nu_0$ quickly decreases $\tau$ so as to
make escape more probable. Furthermore, both profiles imply a net inward
velocity of the gas; for infalling gas, red photons are shifted into resonance,
while blue photons escape even more easily, thus enhancing the blue peak and
diminishing the red.

Fitting a ``Neufeld profile'' to the observed spectra can give us an idea of
the intrinsic characteristica of the system. Unfortunately, there is a
degeneracy in that the profile is dependent only on the parameter $a\tau_0$,
Assuming some temperature, e.g.~$10^4$ K, representative of most of the
Ly$\alpha$
emitting gas, one could in principle deduce the equivalent column density
$N_{\textrm{{\scriptsize H}{\tiny I}}}$.
Alternatively, if $N_{\textrm{{\scriptsize H}{\tiny I}}}$ is 
obtainable due to, e.g., the presence of a background quasar, constraints can
be put on the temperature.

A bulk rotational motion of the gas will also alter the profile.
Since in fact a full spectrum is obtained for every pixel element, this effect
can be studied through long-slit spectroscopy.

\section{Discussion and Conclusion}
\label{sec:disc}
The present MC calculations do not include the effect of dust. 
Effectively, dust will act as a
photon sink and an extra scattering possibility. Since, on average, each
photon scatters $\sim 4 \times 10^{8}$ times and travel a distance of
$\sim 40$ kpc before escaping (determined from a
non-accelerated run of the code with a few
$10^3$ photons), approximately half of which is in the high
density ($n_{\textrm{{\scriptsize H}{\tiny I}}} \gtrsim 0.1$
cm$^{-3}$), cooler ($T \sim 10^4$ K) regions,
even a small amount of dust may be expected to be capable of causing a
significant decrease
in the observed intensity. However, it is not clear to which extend the dust
will affect the observations. If the medium is clumpy, the ratio of Ly$\alpha$
to continuum radiation may in fact be increased \citep{neu91,han06}. Moreover,
\citet{tas06} argued that dust acting as catalysator for hydrogen molecules
will lower $n_{\textrm{{\scriptsize H}{\tiny I}}}$, making the
medium more transparent. We will study these effects in future work,
implementing a realistic model of the dust based on the 8 different metal
species kept track of by the cosmological simulation.

As a very recent improvement of the cosmological simulation, a post-processed
RT scheme of UV radiation from star-forming regions was developed by
\citet{raz06}. They found that up to 5--10\% of the ionizing photons escape these
regions at $z \sim 3.6$. However, a very preliminar analysis indicates that
the results presented in this work are not significantly changed by the
inclusion of H-ionizing photon RT. The quantative effects of this on 
Ly$\alpha$ RT will also be discussed in a forthcoming paper.

The developed Monte Carlo code has reproduced qualitatively
\emph{and} quantitatively the observation that young galaxies often appear
significantly more extended on the sky in Ly$\alpha$ than in the optical.
Furthermore, we investigated the impact of the viewing angle on the
observed surface brightness. Future simulations of a large statistical sample
of galaxies, taking properly into account dust and H-ionizing UV photon 
radiative transfer, will allow us to learn more about the enigmatic Ly$\alpha$
emitters.
%
%
\acknowledgments
The authors are very grateful to J.~Fynbo and T.~Haugb{\o}lle for beneficial
discussions, to A.~Tasitsiomi for quick response to questions, and to
J.~Schaye for useful comments. We also thank the anonymous referee for
constructive remarks. The simulations were performed
on the SGI Itanium II facility provided by DCSC. The Dark Cosmology Centre is
funded by the DNRF.

\end{document}